`

# Trophic Cascades and Habitat Suitability in Udanti Sitnadi Tiger Reserve: Impacts of Prey Depletion and Climate Change on Predator-Prey Dynamics

Krishnendu Basak[1,*], Chiranjib Chaudhuri[2,*], M Suraj[1], Moiz Ahmed[1]

1. *Department of Wildlife Conservation, Nova Nature Welfare Society, H. No. 36/337, Choti Masjid, Byron Bazar, Raipur, Chhattisgarh 492001, India. Correspondence: Tel. +91 9836971670. Email:bastiger08@gmail.com (Basak)*
   *Email: mat.suraj@gmail.com (Suraj)*
   *Email:novanature2004@gmail.com (Ahmed)*
2. *AeroLine International. 181 Littleton Road, Unit 245, Chelmsford, massachusetts, United States 01824. Correspondence: Email: chiranjib@aerolineinternational.com (Chaudhuri)*

This study investigates the trophic cascades and habitat suitability in Udanti Sitnadi Tiger Reserve (USTR), highlighting the roles of apex predators, subordinate predators, and prey species in maintaining ecosystem balance. Using the Trophic Species Distribution Model (SDM), we explored prey-predator interactions and habitat suitability, revealing that tigers, due to prey depletion, increasingly rely on cattle, while leopards adapt by preying on smaller species. The study emphasizes the need for prey augmentation and habitat restoration to support apex predators. Additionally, climate change projections for 2021-2040 and 2081-2100 under CMIP6 scenarios SSP245 and SSP585 indicate significant regional habitat shifts, necessitating adaptive management strategies. Kuladighat is projected to face habitat contraction, while Sitanadi may experience habitat expansion. Effective conservation efforts such as habitat restoration, prey augmentation and predator recovery are the most important steps needed to maintain the purpose of a Tiger reserve and conservation potential of Chhattisgarh-Odisha Tiger Conservation Unit (TCU). To achieve these dynamics, focusing on community participation, anti-poaching measures, and scientific recommendations are the most crucial components to focus on. This comprehensive analysis underscores the critical role of targeted conservation activities in prey-depleted landscapes to ensure the long-term survival of tigers and the overall health of forest ecosystems, enhancing biodiversity and mitigating human-wildlife conflicts in USTR.

# BACKGROUND

A trophic cascade is described as a process by which a perturbation propagates either up or down a food web with alternating negative and positive effects at different successive levels (Terborgh et al. 2006). Large carnivores are categorized by their large body size and for being apex predators, placed high in the trophic ladder (Edwards 2014). As top predators, they can inhibit the eruption of herbivore and subordinate predator populations in ecosystems, an effect that cascades throughout ecological communities and promotes biodiversity (Wallach et al. 2015). The effect of such disappearance of apex predators proceeds downwards successively across the lower trophic levels, resulting in population increases of mid-sized predators i.e. mesopredator release (Crooks and Soul 1999; Prugh et al. 2009) or a higher abundance of subordinate predators, which may affect herbivores and local vegetation in various ways

As various ecosystems globally are facing heavy extirpation of apex predator populations due to habitat loss and persecution from humans, means the removal of their impacts down the line towards subordinate prey species and lower trophic levels, resulting in mesopredator release' therefore populations of subordinate predators have often increased in those habitats (Estes et al. 2011; Ripple et al. 2014). The mesopredator-release hypothesis predicts that, if present, apex or top predators will dominate their subordinate trophic levels, whereas if removed, the successive counterparts will be' released' from dominance and may increase in their numbers (Prugh et al. 2009).

Studies in North America, Europe, and Australia indicated wolves, lynx, and dingoes as top predators controlled subordinate predators like coyotes and foxes, and further checked their impact on lower successive levels (Elmhagen and Rushton 2007; Berger et al. 2008; Letnic et al. 2012). Tiger is the most iconic predator species in Asia, although they suffered serious population decline from anthropogenic pressures: prey depletion by human hunting, elimination of tigers for conflict mitigation or hunted for trading their body parts, and habitat loss or degradation. In spite of conservation efforts over 50 years, wild tigers now occupy <7% of their historic range. Reproducing tiger populations survive in <1% of the ~1.6 million km$^2$ potential habitat (Karanth et al. 2020).

In India, the tiger is identified as a large predator that occupies the top role in the food chain in India. They play a crucial role by exerting regulatory pressure on subordinate predators and herbivore populations, thereby regulating and maintaining the balance of forest ecosystems. Removal or local extinction of such predators may alter the stability of the ecosystem and bring considerable adverse changes. Securing tigers thus safeguards micro-niches in the forest ecosystem, which conserves life forms at the smallest levels, ensuring water and climate security as well (Jhala et al. 2020). Moreover, as a highly threatened and flagship species, the tiger absorbs a continuous flow of funding, which in turn supports a wide range of conservation benefits in India. As a top predator tiger is successfully surviving in the country as compared to other tiger-ranging countries in Asia. In the last few decades, studies and monitoring programs aimed at large carnivore ecology in India has revealed that the tiger population is largely stable and increasing in the country. The project Tiger, started in 1972 with nine tiger reserves (~18,278 km$^2$), has now extended to 53 tiger reserves (~75,796 km2) and successfully engaged about 2.23% of the geographical area of India that supports conservation of representative ecosystems and biodiversity therein (Qureshi et al. 2023). This rising population of tigers is sharing their space and resources with other co-predators. Leopards among these co-predators are majorly found to co-occur with tigers in various landscapes across India. Based on various ecological and administrative factors in all over India, the abundance of tigers is not evenly distributed; rather, many of the tiger reserves have insignificant number of tigers or no tigers at all, leaving the habitats for subordinate predators to thrive or decline in overall trophic levels are prominent.

This study was conducted in the Udanti Sitnadi Tiger Reserve (USTR), which spreads over 1,843 km$^2$ of Gariyaband and Dhamtari Districts of the State of Chhattisgarh. This forest is famous for the tiger *Panthera tigris* and the Asiatic wild buffalo, *Bubalus arnee*. The flora in USTR consists of southern tropical dry deciduous mixed forests and tropical dry peninsular sal forests. From the starting it is experiencing the issues of political unrest that restricted all the ecological monitoring activities and paid the toll in terms of wildlife management (Stripes 2011; Putul 2021; Noronha 2022). Previous publications revealed that USTR had only one or two tigers from 2016–2022 (Jhala et al. 2020; Qureshi et al. 2023; Basak et al. 2023) in the entire tiger reserve. Apart from tigers, this landscape has a small population of wild buffalo, an isolated population of mouse deer, and forest areas with recently invaded transient herds of elephants. Moreover, it houses 246 various species of wetland and terrestrial birds (Bharos et al. 2018), all of these species made this landscape very diverse, and conservation worthy. The area's hilly topography is intercepted by plain strips which together play an important role in connectivity of the Chhattisgarh-Odisha Tiger Conservation Unit (TCU). In the east, the tiger reserve is contiguous with the proposed Sonabeda Tiger Reserve in Odisha, forming the Udanti-Sitanadi-Sonabeda Landscape which spreads over 3000 sq km. In the west, the tiger reserve is connected to Indravati Tiger Reserve in the Bastar region, and in the north, it is connected to Dhamtari and Gariyaband Forest Divisions and further to Barnawapara WS in Mahasamund District. Therefore, this TCU has the potential of significant future in wildlife conservation (Qureshi et al. 2023; Basak et al. 2023).

A few previous studies investigated on the status of tiger, co-predators and their prey species in USTR along with their possible threats in the tiger reserve (Basak et al. 2020; Basak et al. 2023) which puts some lights about wildlife conservation in the landscape which was unnoticed for a long period of time because it was considered as red corridor. Studies highlighted that generally, removing of top predators results in releases of subordinate predators into the habitat, but in case of USTR despite the low the population size of tigers, the leopard population did not increase as expected rather found to be plummeted along with their sparse prey population in the presence of higher human activity across the landscape (Basak et al. 2023). Forest dwelling communities living inside and outside the reserve are using USTR as their hunting ground and continue their traditional practice in the areas of the reserve (Basak et al. 2020). These age-old uncontrolled hunting traditions pose serious threats to the wild ungulate populations, consequently affecting the food resources of carnivore populations in the study area. Moreover, USTR was under prolonged political violence and social unrest that acted as potential hindrance for curbing illegal activities in USTR by using existing legal frameworks. In this critical situation of wildlife conservation in USTR, interventions through species recovery plans are needed urgently. Recovery projects need selection of suitable sites that match the biotic and abiotic needs of the focal species under current and future climates (Swenning J-C, et al. 2016). Therefore, at this time we aimed at unfolding the prey-predator interactions and habitat suitability for tiger and leopard by using the trophic SDM (wedSDM, https://github.com/giopogg/webSDM) model (Poggiato et al. 2022) in USTR. The trophic SDM is a statistical distribution model that models the distribution of species by involving known trophic interactions among the species network present in an area. It provides a useful insight from the trophic cascades concept in ecological science, which is practical in conservation and management demands. Careful use of these obtained results from trophicSDM can be highly useful for wildlife managers and decision makers to predict and prioritize areas for species recovery in a prey-depleted landscape like USTR with the assistance of local communities.

The objective of our study is three-fold, (1) **Understanding the Trophic Cascade of USTR**: This objective aims to investigate the cascading effects within the Udanti Sitnadi Tiger Reserve (USTR). By examining the interactions between different trophic levels, we will gain insights into how the presence or absence of apex predators, such as tigers, influences the populations of subordinate predators and prey species, ultimately affecting the entire ecosystem, **(2) Establishing a Trophic Species Distribution Model (SDM)**: The goal here is to develop a trophic SDM that incorporates known trophic interactions among species in USTR. This model will help us understand the intricate connections between tigers, leopards, and their prey. By integrating biotic and abiotic factors, the model aims to provide a detailed analysis of species distributions and habitat suitability, offering valuable insights for conservation efforts and (3) **Formulating Comprehensive Recommendations for Apex Predator Habitat Recovery**: Based on the findings from the trophic SDM and field observations, this objective focuses on developing actionable recommendations for the recovery and management of apex predator habitats in USTR. These recommendations will be grounded in mathematical modeling and empirical data, aiming to enhance habitat suitability, mitigate human-wildlife conflicts, and support the long-term survival of tigers and other co-predators within the reserve.

## STUDY AREA

USTR is spread over 1842.54 km2 of Gariyaband and Dhamtari districts of Chhattisgarh, central India (Fig. 1). It is constituted with Udanti and Sitanadi Wildlife Sanctuaries as cores and Taurenga, Indagaon, and Kulhadighat Ranges as buffers. The topography of the area includes hill ranges with intercepted strips of plains and lies in the basin of the Mahandi River. The forest types are chiefly dry tropical peninsular sal forest and southern tropical dry deciduous mixed forest (Champion and Seth 1968). Sal is dominant, mixed with *Terminalia sp., Anogeissus sp., Pterocarpus sp.,* and bamboo species. Sitandi side includes dry teak forest, dry peninsular sal forest, and north dry mixed deciduous forest, as per Champion and Seth, 1968. Natural teak forests are mostly found in patches on the alluvial soil along the streams and rivers, while teak plantations have been established in other areas (Kanoje, 2008). *Schleichera oleosa, Terminalia arjuna, Terminalia tomentosa, Mangifera indica, Syzigium cumini, Eugenia heyneana, Ficus racemosa, Ficus lacor, Ficus bengalensis,* and *Stereospermum chelonoides*, all of which are characteristic of riverine or riparian areas.

The Tiger is the apex predator in USTR, and other co-predators are the Leopard, Dhole, Indian Grey Wolf, Striped Hyena, and Sloth Bear. Various wild ungulate species are available as prey bases in the tiger reserve, ranging from small-sized Indian Mouse Deer, Four-horned Antelope, and Barking Deer, mid-sized Chital and Wild Boar, to large-sized Gaur, Sambar, and Nilgai. Smaller carnivores include the Jungle Cat, Rusty-spotted Cat, and Golden Jackal. USTR is contiguous with Sonabeda Wildlife Sanctuary (a proposed tiger reserve) in Odisha on the eastern side and forms Udanti-Sitanadi-Sonabeda Landscape. This connectivity has a good future if the entire tiger landscape complex (Chhattisgarh-Odisha Tiger Conservation Unit) can be taken under significant wildlife conservation efforts.

## METHODOLOGY

**Collection of Species-Occurrence Data**

Occurrence data was collected from repeatedly conducted camera trapping surveys in USTR in 2016 to 2017 and in 2018. We conducted camera trap-based surveys under Phase IV tiger monitoring framework, in 2016-17 to obtain captures of large predators in USTR mainly focused on tiger bearing areas or at least where the chances were higher of having photo-captures of co-predators. The ranges were divided into 2x2 sq. km grids and those grids were used for deploying cameras. Overall, 136 camera trap stations were installed in this session, in three different blocks across North Udanti, South Udanti, and Kulhadighat ranges in 2016-17. The next camera trapping session was carried out during All India Tiger Monitoring (AITM) program in 2018, 2 sq km grid size was used for camera trapping. There we covered Arsikanhar, Risgaon, Sitanadi and Kulhadighat ranges by installing 182 cameras. We deployed the cameras as per the results obtained from carnivore sign surveys before camera trapping.

The total sampling duration was 90 days (about 3 months) each for 2017 and 2018 camera trapping sessions, while cameras were operational for 30 days in each block. The Two camera traps were deployed in each location around forest trails based on indirect evidence of wild carnivore utilization, where possibility of photo-capturing them was higher. We deployed each camera at least 4–5 m from the center of each trail to capture full frame pictures of predators. All cameras were placed at knee height (1.5 feet) from the ground level to obtain identifiable animal-flank photographs. Photo-captures of various herbivore and carnivore species obtained from these two camera trapping sessions were arranged to organize the data based on their presence and absence in the study area. We assigned 1 and 0 values for species presence and absence respectively for each camera point within the sampling time frames.

**Collection of Animal-Trap Data**

Hunting by animal traps is now an alarming issue in places where biodiversity and hunting communities co-occur. Wild animals are often scared, suffocated and killed brutally while entrapped in animal traps. The list of species who suffer the traumatic killing in traps may start from a small rodent to an animal as large as elephants. Controlling or reducing such criminal activities needs very tedious and hard efforts from the concerned conservation authorities. Chhattisgarh is no exception in this case. USTR homes primitive hunting communities who use these forests as their traditional hunting ground. With the advancement of the surrounding world gradually these communities also have adapted urban materials to manufacture traps to catch animals instead using bows, arrows and natural materials. This landscape is a mosaic of forests and human dominated areas. Villages in this landscape have mostly tribal populations belonging to Kamars, Baigas, Gonds, Bhunjiyas and miscellaneous tribes who continued their traditional hunting for bush meat consumptions, earning, recreation and sports as well. Therefore, in such condition a suitable plan has been conceptualized to reduce the effect of snare traps and wildlife poaching in the area or to initiate a habit to curb down the wildlife related crimes by involving the communities and the concerned Government authority. To detect animal traps in USTR, Anti-snare walks were conducted in 2021. Frontline forest staffs were trained to detect and destroy various animal traps used to kill various mammalian species that ranges from small rodents to large herbivores like sambar. Overall, 97 beats were walked to uninstall snares and other traps. On average overall search effort was 6.21 km/walk, for USTR.

**GIS-Data Pre-processing**

In our GIS analysis, it was determined that the minimum distance between camera traps was 200 meters. Therefore, for our study, we utilized a 100-meter grid, which is half of the estimated minimum

`

distance, to ensure finer spatial resolution and more detailed extraction of zonal statistics. These variables encompass a wide range of environmental and geographical attributes, including distances, aspect categories, land cover, and bioclimatic variables.

The distance variables were derived from field data and GIS, providing essential spatial context. The Min_Snare_Distance represents the minimum distance to an animal trap within each 100m grid, with animal trap positions obtained from field surveys. The River Distances, Road Distances, and Village Distances indicate the distance to the nearest river, road, and village, respectively, from each 100m grid. These layers were extracted from OpenStreetMap (OSM), and distance is calculated using GIS techniques in R, providing critical spatial relationships within the study area.

Aspect and slope categories were derived from Copernicus 30m Digital Elevation Models (DEM) and processed using GIS techniques. The DEM used is the Copernicus digital elevation model. Aspect categories include North-facing slopes, East-facing slopes, South-facing slopes, and West-facing slopes. These variables represent the fraction of each aspect category within the grid cells. Similarly, slope categories were divided into gentle slopes, <5 degrees, moderate slopes, 5-15 degrees, steep slopes, 15-30 degrees, and very steep slopes,>30 degrees. These categorizations as grid fractions help in understanding the terrain's variability and its potential impact on the study outcomes. Additionally, we used the mean elevation within each grid cell as an input variable, derived from the DEM.

Land Use and Land Cover (LULC) categories were sourced from ESRI and provide critical insights into the landscape composition. These layers represent forests, rangelands and croplands. These LULC categories, represented as grid fractions, help in understanding the distribution of different land cover types within the study area.

The bioclimatic variables were obtained from WorldClim data and include a comprehensive set of climatic attributes. These variables are Annual Mean Temperature, Mean Diurnal Range, Isothermality, Temperature Seasonality Max Temperature of Warmest Month, Min Temperature of Coldest Month, Temperature Annual Range, Mean Temperature of Wettest Quarter, Mean Temperature of Warmest Quarter, Mean Temperature of Coldest Quarter, Annual Precipitation, Precipitation of Wettest Month, Precipitation of Driest Month, Precipitation Seasonality, Precipitation of Wettest Quarter, Precipitation of Driest Quarter, and Precipitation of Warmest Quarter. These variables provide a detailed climatic profile of the study area, which is essential for understanding environmental influences on the analyzed phenomena.

By compiling these variables, we ensured a comprehensive representation of the study area's physical, climatic, and geographical characteristics. The integration of field data, GIS-derived metrics, DEM, ESRI land cover categories, and WorldClim bioclimatic variables (Table 1) facilitates a robust and multidimensional analysis of our research questions. To improve model performance, we applied a quantile transform to the input variables with a target normal distribution. This transformation helps normalize the data distribution, enhancing the efficiency and accuracy of the predictive models used in our study.

`

**Trophic Model**

       The Trophic Species Distribution Model (trophicSDM) (Poggiato et al, 2022) utilized in this study provides a sophisticated framework for understanding predator-prey interactions within the Udanti-Sitanadi Tiger Reserve (USTR). By integrating trophic relationships alongside abiotic factors, such as climate, terrain, land cover, and distance from anthropogenic layers, this model offers a detailed analysis of species distributions and habitat suitability. TrophicSDM sheds light on the ecological consequences of varying apex predator populations, such as tigers, and their influence on subordinate predators, such as leopards, and their prey species. This approach is especially pertinent in ecosystems like USTR, where apex predator numbers are critically low, allowing for the examination of ecological changes. In USTR, trophicSDM identified essential interactions between tigers, leopards, and their prey, highlighting the necessity of maintaining balanced trophic dynamics and considering both biotic and abiotic factors for effective conservation

       The hypothesized trophic connectivity diagram (Figure 2) shows the interactions between prey and their predator species in UTSR, specifically with tigers and leopards as the primary predators. Tigers are connected to a range of prey species, including Sambar, Nilgai, Indian Gaur, Wild Pig, Spotted Deer, and cattle; tigers are majorly inclined towards the large to middle sized prey species (Karanth & Sunquist 1995, Hayward et al. 2012, Basak et al. 2018, Basak et al. 2020). Tigers' preference for larger prey provides insights their requirement of substantial biomass, which aligns with their status as apex predators requiring significant energy intake. Tigers' inclusion of livestock in their diet can be attributed to the availability of higher biomass cattle, especially when wild prey is scarce. Whereas, leopards are connected to Northern Plains Langur, Indian Hare, Chousingha, Wild Pig, Spotted Deer, and cattle. Leopards are known for their adaptability to changing environment, often by widening their food choices as per the availability of prey species that includes varied prey sizes as well (Eisenberg & Lockhart 1972; Santiapillai et al. 1982; Johnsingh 1983; Rabinowitz 1989; Seidensticker et al. 1990; Bailey 1993; Karanth and Sunquist 1995; Daniel 1996; Edgaonkar and Chellam 1998; Sankar and Johnsingh 2002; Qureshi and Edgaonkar 2006; Edgaonkar 2008; Mondal et al. 2011; Sidhu et al. 2017). This flexibility in terms allows them to survive in diverse environments, including areas close to human settlements where livestock might be more accessible. Leopards' prey includes both smaller animals, like Indian Hare and Northern Plains Langur, and larger prey, such as cattle and Spotted Deer, showcasing their adaptability and opportunistic feeding behaviour (Basak et al. 2020).

       The Figure 2 also highlights the differences in biomass and size preferences between tigers and leopards. Tigers prefer larger prey that meets their high energy demands, while leopards exhibit a broader dietary range, adapting to prey availability and habitat conditions. The model has derived trophic relations between the predator and prey species of USTR and enhanced understanding how the bottom-up effect can impact the predators in such landscape where availability of both large and middle-sized prey species is shaping the large predator interactions and population dynamics. Trophic

`

connections were derived at different confidence levels: 90% (a), 80% (b), and 70% (c). It is predicted that at 90% confidence, the connections will be conservative and show only significant relations. Whereas 80% and then 70% will flexible the limit and can exhibit the wider range of interactions that can be suitable for tigers in USTR, if prey sample increases.

**Model Setup for Trophic Interactions**

To set up our trophic model, we utilized a Bayesian framework implemented in the stan_glm function. Our model was designed with a binomial output using a logit-link function to accurately capture the probabilistic nature of abiotic and biotic interactions. We ran two Markov Chain Monte Carlo (MCMC) chains, each with a total of 2000 samples. To ensure the convergence and stability of our estimates, we specified a burn-in period of 1000 samples for each chain, discarding these initial samples to mitigate the influence of the starting values.

**Justification for using Trophic Model**

The justification for using the trophic model over the non-trophic model for tigers and leopards is compelling based on the provided statistics. For tigers, the trophic model exhibits a higher AUC (0.84 vs. 0.79) and TSS (0.66 vs. 0.51) based on the mean from a 5-fold cross-validation experiment, indicating superior predictive performance. Although the non-trophic model shows slightly higher AUC (0.69 vs. 0.66) and TSS (0.33 vs. 0.27) for leopards, the overall fit statistics favor the trophic model, which has a lower AIC (4442.54 vs. 4449.92) and a higher log-likelihood (−1817.27 vs. −1828.96), indicating a better fit to the data. The poorer AUC for leopards could be due to their behavior as habitat generalists, thriving in diverse environments and making their presence more challenging to predict. In summary, the trophic model demonstrates better predictive power and fitness, particularly for tigers. By capturing crucial ecological interactions, it offers a more accurate and comprehensive representation of their natural habitats and behaviors, making it a valuable tool for conservation efforts.

**Habitat Experiments**

We estimated the optimal threshold for each of the 200 MCMC inference samples using the Youden method, which identifies the threshold that maximizes the sum of sensitivity and specificity, thereby balancing true positive and true negative rates. Then, using a 90% confidence level of prediction, we set the 10th percentile of the sample optimum thresholds as a fixed threshold for experiments. Subsequently, we classified any region with a habitat suitability value higher than this

`

threshold as Habitat, and values lower than this threshold as Not Suitable. For the subsequent habitat change experiment, we compared the realized habitat with the post-experiment simulated habitat. Grids that are not habitat in both scenarios are classified as NO (no habitat). Habitats that remain unchanged are classified as NC (no change). Areas where habitat expands are classified as RE (range expansion), and areas where habitat contracts are classified as RC (range contraction).

**Climate Change Experiment Setup**

For our climate change impact assessment on habitat suitability in the Udanti Sitanadi Tiger Reserve, we utilized habitat projections for the periods 2021-2040 and 2081-2100 under scenarios from the Coupled Model Intercomparison Project Phase 6 (CMIP6), specifically SSP245 (low-emission scenario) and SSP585 (high-emission scenario). We employed four climate models: CMCC-ESM2, GISS-E2, HadGEM3-GC3, and UKESM1. The methodology involved swapping the bioclimatic variables from the abiotic layers with their equivalent bias-corrected bioclimatic variables for each model-scenario-time slice combination.

# RESULTS

**ROC and habitat predictions**

The ROC (Receiver Operating Characteristic) curves (Figure 3) provide a visual representation of the model's performance, with sensitivity (True Positive Rate) plotted against 1 - specificity (False Positive Rate). The red line indicates the median ROC curve, representing the average performance across all 200 MCMC (Markov Chain Monte Carlo) inference samples. The shaded blue area around the median ROC curve is the 95% uncertainty envelope, reflecting the variability in model performance across different MCMC samples and showing the range within which the true ROC curve is likely to lay 95% of the time.

The tiger model's ROC curve (Figure 3a) lies significantly above the black diagonal line of random prediction, demonstrating high sensitivity and specificity. The close proximity of the median curve to the upper left corner indicates excellent model performance, and the narrow uncertainty envelope underscores the reliability and robustness of the model's predictions.

`

Compared to the tiger model, the leopard model's ROC curve (Figure 3b) is closer to the black diagonal line of random prediction, indicating lower sensitivity and specificity. The broader 95% uncertainty envelope suggests greater variability in the model's predictions. While the median ROC curve does rise above the random prediction line, the performance is less pronounced, reflecting the challenges in accurately predicting leopard presence due to their behavior as habitat generalists and other factors influencing their distribution.

**Predictor Significance Analysis of Habitat Suitability in USTR**

The habitat suitability analysis for various species in the Udanti Sitanadi Tiger Reserve provides insights into the critical abiotic and biotic factors influencing their distribution (Table 2). This discussion aims to elucidate these factors to develop effective conservation and management strategies within the reserve.

**Tiger Habitat Suitability**

In our trophic species distribution model (SDM) for tiger habitat, several variables emerged as significant predictors, each contributing differently to the model. The intercept, with a mean value of -9.45, represents the baseline log-odds of tiger presence when all predictor variables are set to zero. This negative value suggests that, in the absence of other factors, the likelihood of tiger presence is inherently low, highlighting the importance of the additional variables in predicting suitable tiger habitats. We have mainly observed tigers in Kuladighat and Phase IV Block 2, with detections at 22 sites, and a total count of 45 observations. In our analysis, 255.88 sq km is found to be suitable for tiger habitat (Figure 4a).

One of the most significant variables is the minimum distance to the nearest snare (Min_Snare_Distance), which has a mean coefficient of -1.48. This indicates that tigers are more likely to be found near snares, suggesting higher habitat suitability in these areas, likely due to higher prey abundance. However, this proximity increases the risk of animals being caught in traps, underscoring the urgent need for anti-poaching measures. Addressing this issue is crucial for protecting both prey and predator species within the reserve.

Aspect category 2 (ASPECT_cat_2), representing East-facing slopes, has a positive mean coefficient of 0.36. This suggests that tigers are more likely to be found in habitats with these slope characteristics. Additionally, steep slopes categorized under SLOPE_cat_4 (very steep slopes, >30 degrees) also positively influence tiger presence, with a mean coefficient of 0.80. These findings indicate that tigers may prefer certain topographical features, possibly due to the cover and hunting advantages they provide.

`

Among the bioclimatic variables, wc2.1_30s_bio_12 (Annual Precipitation) and wc2.1_30s_bio_13 (Precipitation of Wettest Month) shows contrasting effects. Annual precipitation has a negative mean coefficient of -5.49, suggesting that higher annual precipitation levels are associated with lower tiger presence. In contrast, precipitation during the wettest month has a positive mean coefficient of 5.46, indicating that areas with high precipitation in the wettest month may be more favourable for tigers. This contrast highlights the complex relationship between precipitation patterns and tiger habitat suitability.

Lastly, the presence of cattle is a significant positive predictor, with a mean coefficient of 2.60. This suggests despite tigers tend to hunt wild ungulates with large biomass however in critically low abundance of large-sized ungulates in USTR they exploit available cattle population to meet their energy needs in the reserve. This preying on cattle indicates a potential human-tiger conflict in this area.

**Leopard Habitat Suitability**

In our trophic species distribution model (SDM) for leopard habitat, several variables emerged as significant predictors, each contributing differently to the model. The intercept, with a mean value of -2.87, represents the baseline log-odds of leopard presence when all predictor variables are set to zero. This negative value suggests that, in the absence of other factors, the likelihood of leopard presence is inherently low, underscoring the importance of the additional variables in predicting suitable leopard habitats. Unlike tiger we have observed Leopard presence in almost all the subdivisions, with detections at 159 sites, and a total count of 388 observations. In our analysis, 430.17 sq km is found to be suitable for Leopard habitat (Figure 4b).

Aspect categories play a notable role in determining leopard habitat preferences. ASPECT_cat_1 (North-facing slopes) has a positive mean coefficient of 0.15, indicating that leopards are more likely to be found on these slopes. Conversely, ASPECT_cat_4 (West-facing slopes) has a negative mean coefficient of -0.12, suggesting that these slopes are less suitable for leopards. This highlights the importance of specific topographical features in influencing leopard distribution.

Land Use and Land Cover (LULC) categories also significantly impact leopard habitat suitability. ESRI_LULC_cat_11 (forests) has a substantial positive mean coefficient of 6.13, indicating that forested areas are highly favorable for leopards. Similarly, ESRI_LULC_cat_2 (rangelands/croplands) also shows a strong positive influence with a mean coefficient of 6.23. These findings suggest that leopards prefer a mix of forested and open landscapes, which provide both cover and hunting opportunities.

Slope category 3 (SLOPE_cat_3), representing steep slopes (15-30 degrees), has a positive mean coefficient of 0.16. This indicates a slight preference for steeper slopes, which may offer advantages in terms of cover and vantage points for hunting.

Among the bioclimatic variables, wc2.1_30s_bio_11 (Mean Temperature of Coldest Quarter) and wc2.1_30s_bio_12 (Annual Precipitation) both show positive influences on leopard presence, with mean coefficients of 6.36 and 1.28, respectively. This suggests that leopards are more likely to be found in areas with moderate to high precipitation and cooler temperatures during the coldest quarter. Conversely, wc2.1_30s_bio_8 (Mean Temperature of Wettest Quarter) has a negative mean coefficient of -3.89, indicating that higher temperatures during the wettest quarter are less favorable for leopards. Additionally, wc2.1_30s_bio_5 (Max Temperature of Warmest Month) has a positive mean coefficient of 2.28.

Prey availability is another crucial factor influencing leopard habitat suitability. The presence of wild pigs (Wild_Pig) has a positive mean coefficient of 0.54, indicating that areas with higher wild pig populations are more likely to support leopards. Similarly, the presence of Northern Plains Gray Langurs (Northern_Plains_Langur) and Indian Hares (Indian_Hare) positively influences leopard presence, with mean coefficients of 1.05 and 0.80, respectively. These prey species provide essential food resources for leopards, enhancing habitat suitability.

**Habitat Experiments**

The Figure 5 illustrates significant trophic connections between predators and prey at UTSR across three confidence levels: (a) 90%, (b) 80%, and (c) 70%. At 90% confidence (Figure 5a), the connections are conservative, showing only the most significant interactions, such as tigers connected to cattle, and leopards to Northern Plains Langur and Wild Pig. As the confidence level decreases to 80% (Figure 5b) and then to 70% (Figure 5c), the tiger exhibits increasing prey width, adding connections to Sambar and Spotted Deer. This suggests that due to a lower tiger population and the higher biomass availability of cattle, tigers have shifted their diet towards cattle. However, tigers have a natural dietary preference for Sambar and deer, which are not available in abundance within their habitat area due to low populations. In contrast, the leopard's interactions remain largely unchanged across the different confidence levels, indicating a consistent prey base. This expansion of prey connections for tigers at lower significance levels highlights the complexity and broader scope of their trophic interactions compared to leopards.

In this experiment, we aimed to compare the realized niche with the fundamental niche of tigers (Figure 6a) by manipulating prey availability. Specifically, we excluded cattle from the tiger's diet and made Sambar and deer available throughout the study area. The realized niche represents the actual conditions and resources a species utilizes in the presence of biotic interactions, such as competition and predation, while the fundamental niche encompasses the potential range of conditions and resources a species could theoretically use without such interactions. Tigers have increasingly relied upon cattle due to the low wild ungulate population and the high biomass of available livestock. By ensuring the widespread availability of their preferred natural prey, such as Sambar and deer, and removing the availability of livestock, we observed changes in the tigers' habitat use and distribution (Figure 6b). This adjustment was made to understand the tiger's natural dietary preferences and habitat requirements better, free from the constraints of current prey availability. In this experiment, we observed the following positive habitat metrics: 1512.37 sq. km of No Occurrence (NO), a minimal 1.08 sq. km of Habitat Contraction (RC), a substantial 194.45 sq. km of Habitat Expansion (RE), and a stable 254.8 sq. km of No Change (NC). These results provide a solid reference point for understanding the beneficial impact of prey manipulation on tiger habitat dynamics, highlighting the potential for improved habitat conditions when tigers' natural prey preferences are supported.

**Climate Change Scenario Analysis**

The habitat projections for 2021-2040 under scenarios CMIP6 SSP245 and SSP585, using models CMCC-ESM2, GISS-E2, HadGEM3-GC3, and UKSM1 reveal significant changes across the Kuladighat and Sitanadi regions (Figure 7). The analysis indicates considerable habitat changes in both scenarios (Table 3), with more pronounced impacts under the high-emission SSP585 scenario.

For period of 2021-2040, in the SSP245 scenario (Figure 7a), the CMCC-ESM2 model shows habitat expansion in Sitanadi (RE: 1519 sq. km) with no contraction. However, under the SSP585 scenario (Figure 7b), this model projects small habitat expansions in Sitanadi (RE: 273 sq. km) and range contraction in Kuladighat (RC: 238 sq. km). In the SSP245 scenario (Figure 7c), the GISS-E2 model highlights new habitat expansions in Sitanadi (RE: 1339 sq. km) with negligible range contraction (RC: 7 sq. km). Under the SSP585 scenario (Figure 7d), the GISS-E2 model shows the range expansion in Sitanadi (RE: 173 sqkm) and range contraction Kuladighat (RC: 206 sq. km). In the SSP245 scenario (Figure 7e), the HadGEM3-GC3 model predicts habitat extinction in Kuladighat (RC: 230 sq. km) and shifting of habitat to Sitanadi (140 sq. km). Under the SSP585 scenario (Figure 7f), this model shows significant habitat contractions in Kuladighat (RC: 157 sq. km) and habitat expansion in Sitanadi (RE: 529 sq. km). In the both the scenario (Figure 7g and Figure 7h), the UKSM1 model shows almost similar patterns of change as HadGEM3 with habitat extinction in SSP245 scenario (RC: 247 sqkm, RE: 6 sq. km) and under SSP585 habitat shifting from Kuladighat to Sitanadi (RE: 253 sq. km, RC: 222 sq. km).

For the period 2081-2100, the CMCC-ESM2 model under SSP245 (Figure 8a) continues to predict complete habitat extinction (RC: 254 sqkm). Under the SSP585 scenario (Figure 8b), this model projects vast expansion of habitat in Sitanadi (RE: 1523 sqkm) and negligible habitat contraction in Kuladighat (RC: 20 sq. km). The GISS-E2 model under SSP245 (Figure 8c) shows habitat expansion significantly in Sitanadi (RE: 1306 sq. km) and no discernable habitat contraction (RC: 1 sq. km). However, under the SSP585 scenario (Figure 8d), the GISS-E2 model is showing habitat extinction (RC: 254 sq. km). The HadGEM3-GC3 model under SSP245 (Figure 8e) is showing habitat extinction (RC: 255 sq. km). Under the SSP585 scenario (Figure 8f), this model shows the shifting of habitat from Kuladighat to Sitanadi (RC: 177 sq. km, RE: 1003 sq. km). The UKSM1 model again follows the

`

HadnGEM3 model. In SSP245 (Figure 8g) it is showing habitat extinction (RC: 255 sq. km) and in SS585 (Figure 8h) it showing habitat shifting from Kuladighat to Sitanadi (RC: 195 sq. km, RE: 1135).

## DISCUSSSION

Trophic interaction plays a significant role in maintaining healthy predators and their prey population dynamics in diverse ecosystems that simultaneously sustain the ecological balance of natural habitats. Apex predators majorly occupy high trophic levels; their presence may regulate other predators and prey species at lower levels through trophic cascades (Ripple et al. 2014). Therefore, in the absence of predators, ecosystem functionality may face various risks (Ripple et al. 2014). In the case of African lions, their density is strongly correlated to prey density (Van Orsdol et al. 1985) and thus they faced decrease in response to prey depletion (Vinks et al. 2021) in various areas of African continent. The survival rates and population densities of subordinate competitors such as the African wild dog and cheetah are less tightly correlated with prey density but are negatively correlated with the density of dominant competitors (Creel and Creel 1996; Kelly et al. 1998; Mills and Biggs, 1993; Mills and Gorman 1997; Swanson et al. 2014). Contrary to a long-term study in the Greater Kafue Ecosystem (GKF), Zambia found a decline in the large predator population when the prey population was comparatively low and the and the African wild dog population was comparatively low, and when the lion population of the area was highly declined, as were the prey populations as well. It was expected that African wild dogs would exploit the opportunity and increase the population size at that time, but reality reflected the opposite scenario, where the African wild dog population lived with a comparatively lower population density (Goodheart et al. 2021).

In the Indian context, large predators like tigers exert competitive pressure on other subordinates or co-predators, which in turn may positively influence the prey populations. Therefore, in tiger-ranging areas, studies have observed that lowering the population of top predators like tigers leads to an increase in the subordinate predators like leopards. And reversing the situation by reintroducing tigers to its past range, may significantly lower the leopard population by releasing spatial and dietary competition from tigers (Harihar et al. 2011; Mondal et al. 2013). The previous studies conducted in USTR reflected that the area nearly lost its tiger population, basically surviving by one or two individuals (Qureshi et al. 2022; Basak et al. 2023), and the leopard population was expected to flourish in the absence of the apex predator, but leopards were found to survive with a comparatively low population density of 1.56 ± 0.36 SE/100 km2. Subsequently, the prey species, especially the ungulate species of every size, exhibited the same trend; the derived ungulate density was only 8.46 ± 2.1 SE individuals/km$^2$ (Basak et al. 2023). Prey depletion leads to the decline of large carnivores globally and is usually a very prominent issue in USTR as well. This bottom-up effect was prominent in the trophic experiment using Trophic SDM to check the trophic interaction between the available prey and predators. In this study, the predicted trophic connections at 90% confidence highlighted that tigers only have interactions with cattle and leopards have significant interactions with northern plains gray langur, Indian Hare, and wild pig, but by lowering the confidence level to 80% and 70% with much more flexible options, the model indicated connections between tigers and sambar and spotted deer, whereas leopards remained more or less unchanged and maintained the relationship with northern plains gray langur, wild pig, and Indian hare. Overall, the trophic interaction revealed that in a situation where the prey population is highly plummeted, leopards have expanded their diet width and exploited medium- and small-sized prey like Indian hare, northern plains gray langur, and wild pig, whereas in the absence of large herbivores with larger biomass, which is highly suitable for tigers to maximize their energy intake, they shifted towards free-ranging domestic cattle available in the forest villages or hamlets to gain the energy requirements. Therefore, increasing the sample size for various prey can change the scenario of predator-prey interactions. Thus, recruiting various prey species, especially large and medium-sized ungulates like the gaur, sambar, wild pig, and spotted deer for tigers, is highly crucial for the long-term survival of the species with other co-predators.

Apart from trophic interactions among prey and predator species in USTR, this study found prey-mediated habitat utilization and predator distribution in the landscape. The spatial utilization of a large predator reflects the interactions between various biotic and abiotic factors and bioclimatic parameters, human influences, the presence of competitors, and access to water (Bailey 1993; Marker and Dickman 2005; Vanak et al. 2013; Snider et al. 2021). Central India and the Eastern Ghats tiger landscape complex comparably have the highest tiger population and remarkable spatial occupancy in India, providing due to its good quality habitat and corridor connectivity among the tiger reserves and other forested areas. This tiger landscape complex spread across the sates of Rajasthan, Maharashtra, Madhya Pradesh, Chhattisgarh, Jharkhand, and Odisha. Additionally, it includes the areas from Eastern Ghats in Telangana, Andhra Pradesh, and Odisha, and certain parts of the Northern Western Ghats (Sahyadri) in Maharashtra to maintain the characteristic integrity of the landscape complex (Qureshi et al. 2022). Despite having such quality of habitats that holds highest number of tigers, there are areas like Udanti-Sitanadi Tiger Reserve where quality or suitable habitats were found very low where tigers can thrive, and the population was observed surviving with much lower-than-expected viable population. The 1843 sq. km spanned tiger reserve has only 13% habitat to offer tigers to thrive but thereto severely depleted prey base pushed tigers to survive on domestic cattle. Even suitability experiments predicted only 23% suitable habitat for highly adaptable big cat like leopard in USTR. Leopards occupy a prominent position in the trophic pyramid alongside tigers, exhibiting adaptability in habitat and dietary preferences, and playing a vital role as top predators in a wide array of landscapes across India (Qureshi et al. 2024). Therefore, we considered analysis of leopards' habitat in USTR as well. Attempts revealed that leopards were also observed to survive in low density and shifted to small sized prey species like langur, Indian hare to medium sized prey like wild pig and opportunistic killing of available cattle population in the reserve. Whereas spatial occurrence of animal traps in the reserve indicating heavy depletion of animal by poaching in the reserve which has role to regulate prey and predator population across the time. Thus, protection measures and effective interventions are needed to be implemented to curb the challenges of human dependence in USTR.

Tiger habitat consists of areas with cover for hunting and raising cubs, sufficient availability of prey biomass, preferentially medium- and large-sized ungulates, and freedom from persecution by humans, the tiger's main competitor (Gittleman and Harvey 1982; Karanth and Sunquist 1992; Smith 1993; Miquelle et al. 1999a; Darimont et al. 2023). If we keep aside the anthropogenic parameters, climate change predominantly shapes tigers' habitat in the wild (Cooper et al. 2016). Climate change is globally altering natural ecosystems, habitats and thus niches and their nature of diversity (Bellard et al. 2012). In future climate change together with human activities will cause severe habitat degradation, biodiversity loss, extinction of species and their crucial habitats in places and can potentially push the apex predators like tiger towards extinction.

In this present study area, the detailed analysis of climate change time slices highlights the differential impacts on Kuladighat and Sitanadi areas. The projected habitat changes for 2021-2040 (Figure 7) and 2081-2100 (Figure 8) under CMIP6 scenarios SSP245 and SSP585 using CMCC-ESM2, GISS-E2, HadGEM3-GC3, and UKSM1 models reveal significant regional differences in habitat dynamics in USTR. Kuladighat region, which is the remaining tiger habitat in USTR faces substantial habitat changes, while Sitanadi region where tiger is found to be absent in last decade, is projected to experience expansions and new occurrences. The hilly Kulhadighat, with its less dense human population, appears more vulnerable to habitat changes, even can experience habitat extinction as models predicted, while the plains and lower hills of Sitanadi, with higher human population density, are likely to see increased habitat suitability.

This dichotomy of habitat gains and loss in tiger reserve areas necessitates a dual approach in conservation strategies. For Kulhadighat, efforts should focus on preserving the remaining habitats and restoration of degraded habitats through targeted conservation programs and mitigation of climate impacts. In contrast, Sitanadi will require strategies to manage the influx of new habitats and potential species migrations. Proactive measures such as creating wildlife corridors and buffer zones can help mitigate human-wildlife conflicts and promote biodiversity resilience. Human activities can make the parameters such as maintaining habitat connectivity and species influx vulnerable in future and thus need to be safeguarded with strong protection measures in the landscape.

Globally, tigers experienced 150 years of decline; aftermath effective potential habitat for the tiger seems to have stabilized at around 16% of its indigenous extent (1.817 million $km^2$). There were 63 Tiger Conservation Landscapes in the world as of $1^{st}$ January 2020; spread over 911,920 km2 shared across ten of the 30 modern countries which once harboured tiger populations. Over the last 20 years, the total area of Tiger Conservation Landscapes (TCLs) declined from 1.025 million km2 in 2001, a range-wide loss of 11%, with the greatest losses in Southeast Asia and southern China. Meanwhile there was significant increase in Tiger Conservation landscapes in India, Nepal, Bhutan, Northern China and South-eastern Russia. Overall, 226 restoration landscapes in these areas can give rise to 50% rise in tiger population (Sanderson et al., 2023). We found USTR is such a landscape where careful integration of conservation strategies along with positive community participation and continuous input of strong scientific recommendations can bring thriving future for tigers, co-predators and their prey species in USTR and obviously in turn revive the functionality of the Chhattisgarh-Odisha Tiger Conservation Unit (TCU) and can potentially act as reservoir of largely biologically diverse life forms. These findings underscore the urgent need for adaptive management and conservation strategies tailored to the specific challenges and opportunities presented by the USTR landscape. USTR falls approximately under 35% of the tiger reserves in the country, which urgently needs efficient and effective planning for habitat restoration, ungulate augmentation, and subsequent tiger reintroduction. This will be highly crucial to ensuring sustainable coexistence between human and natural systems in the face of ongoing climate change.

The detailed analysis of climate change time slices reveals the differential impacts on Kuladighat and Sitanadi. The non-dense population of hilly Kulhadighat appears to be more at risk from habitat change, while the higher-density plains of Sitanadi are expected to see enhanced suitability in the future. This dichotomy calls for two-pronged conservation strategies. At Kuladighat, priority should be given to the conservation of habitats in areas that have been left largely unchanged, with efforts to mitigate climate impacts. On the other hand, strategies to manage and increase new habitats in Sitanadi would potentially helpful for species dispersal. Creating wildlife corridors and buffer zones in advance can reduce human-wildlife conflicts and foster the resilience of biological diversity.

The spatial species distribution model predictions for 2021-2040 (Figure 7) and projections by two CMIP6 models under the scenarios SSP245 and SSP585 for the periods 2081-2100 (Figure 8) reveal significant habitat loss for Kuladighat, whereas expansions and new occurrences are forecasted for Sitanadi. This underlines the necessity to account for both climate change model-scenario uncertainty and inter-model variability. This variation in projection among models must be considered whenever these future projections are used to inform conservation planning.

Our results highlight the importance of implementing adaptive regional management and conservation tools to adequately address local threats while utilizing new opportunities. The future of biodiversity and a sustainable coexistence of human systems with natural systems in the backdrop of continued climate change will largely depend on successful planning and implementation. By accounting for these uncertainties, conservation strategies can be made more robust and resilient, ensuring they remain effective under a range of possible future climate scenarios.

## CONCLUSION

The study assesses trophic cascades and habitat suitability in the Udanti Sitnadi Tiger Reserve (USTR) with a focus on interactions among apex predators, subordinate predators, and prey species using comprehensive analytical tools. The results underscore the importance of tigers and leopards in ecosystem regulation, as well as the negative consequences of depleting prey populations and human activity.

The study uses the Trophic Species Distribution Model (SDM) to provide an understanding of the spatial and trophic dynamics of USTR. The findings highlight the need for focused conservation strategies, including habitat restoration and prey supplementation, to help sustain apex predator populations as well as ecological diversity.

Our climate change projections underscore the need for rapid adaptive management to ameliorate habitat shifts and also maintain the ecosystem resilience of USTR. The successful implementation of these approaches will require collaborative efforts with local communities, government agencies, and conservation organizations.

Together, this study leads to better insights into ecological interactions and a strong framework for designing conservation action plans with priorities that are implemented in prey-depleted landscapes such as USTR. Conservation of these iconic species effectively secures their prey base and the overall health of forests not only contributes to tiger conservation but also provides ecosystem services such as clean drinking water and air quality improvement.

**Acknowledgment:** We express earnest gratitude to Shri Kaushalendra Singh, former PCCF (Wildlife) and Shri Sudhir Agarwal, PCCF (Wildlife), Chhattisgarh. We also convey our gratitude to Mr. K. Murugan, former CCF (Wildlife) for his initiative and continuous support during the project implementation period. We would like to convey our deepest gratitude to Shri O. P. Yadav, Member Secretary Chhattisgarh for his trust in Nova Nature Welfare Society and provided with every possible support. We are always grateful to Chhattisgarh State Forest Department to keep faith on us and providing necessary permission and essential financial support to conduct the study. We would also like to sincerely mention that part of this research was supported by The Habitat Trust, Conservation Hero Grant 2020. We are thankful to Shri B.V. Reddy, Divisional Forest Officer, Shri Nair Vishnu Narendran, Divisional Forest Officer and Shri Ayush Jain, Divisional Forest Officer for their constant support during the field. We also want to thank Shri Sunil Sharma, Sub- Divisional Forest Officer for his directions in the field, without such guidance it might be impossible to collect data from the difficult terrain of Udanti Sitanadi Tiger Reserve. We really appreciate Mr. Amit Kumar for his tedious effort regarding GIS data procurement and pre-processing. We would like to convey our sincere thanks to biologist Mr. Chiranjivi Sinha for his rigorous contribution in the field during the tiger monitoring program. We also thank Mr. Ajaz Ahmed, Mr. Nitesh Sahu, Mr. Om Prakash Nagesh and the entire team from Nova Nature Welfare Society for their contribution to field works and all frontline forest staff from USTR for their support during the study.

**Author's contribution:**

1. **Krishnendu Basak;** substantial contributions to conception and design, acquisition of data, or analysis and interpretation of data; drafting the article or revising it critically for important intellectual content; and final approval of the version to be published.

`

2. **Chiranjib Chaudhuri;** substantial contributions to conception and design, data analysis and interpretation of data. Drafting the article or revising it critically for important intellectual content; and final approval of the version to be published.
3. **M Suraj;** Acquisition and interpretation of data. Drafting the article and final approval of the version to be published.
4. **Moiz Ahmed;** Acquisition and interpretation of data. Drafting the article and final approval of the version to be published.

**Competing interests:** All authors declare that they have no competing interests.

**Availability of data and materials:** Data available upon reasonable request.

**Consent for publication:** All of the contributors for this study are supportive for the publication.

**Ethics approval consent to participate:** Not applicable

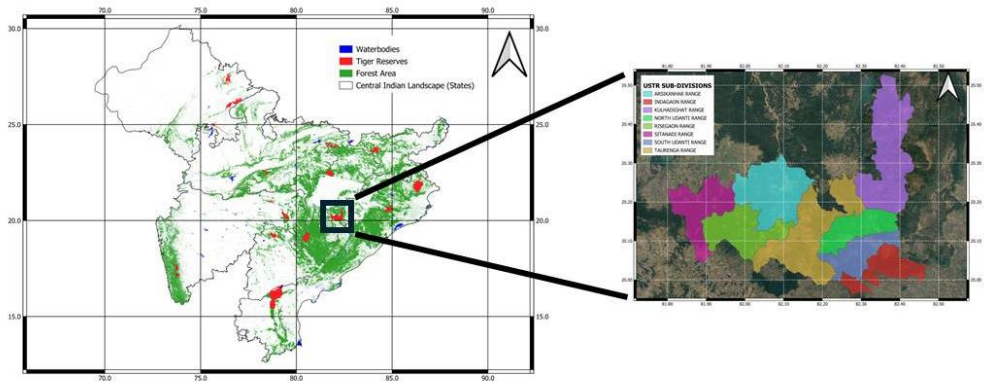

Fig. 1. Map of Central Indian landscape complex (CILC) showing location of present study area Udanti-Sitanadi Tiger Reserve (USTR). (Map Data: Google Satellite Basecamp).

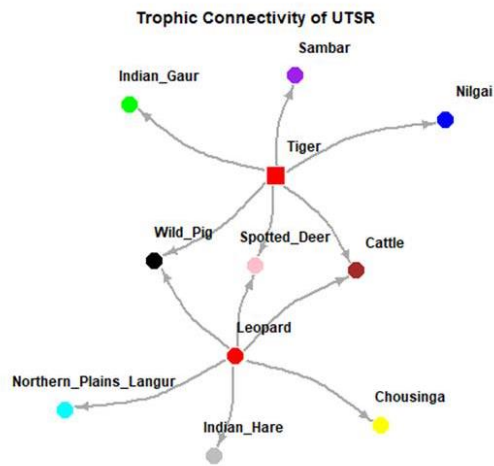

Fig. 2: The Trophic connectivity expected in Udanti-Sitanadi Tiger Reserve (USTR), Chhattisgarh, Central India.

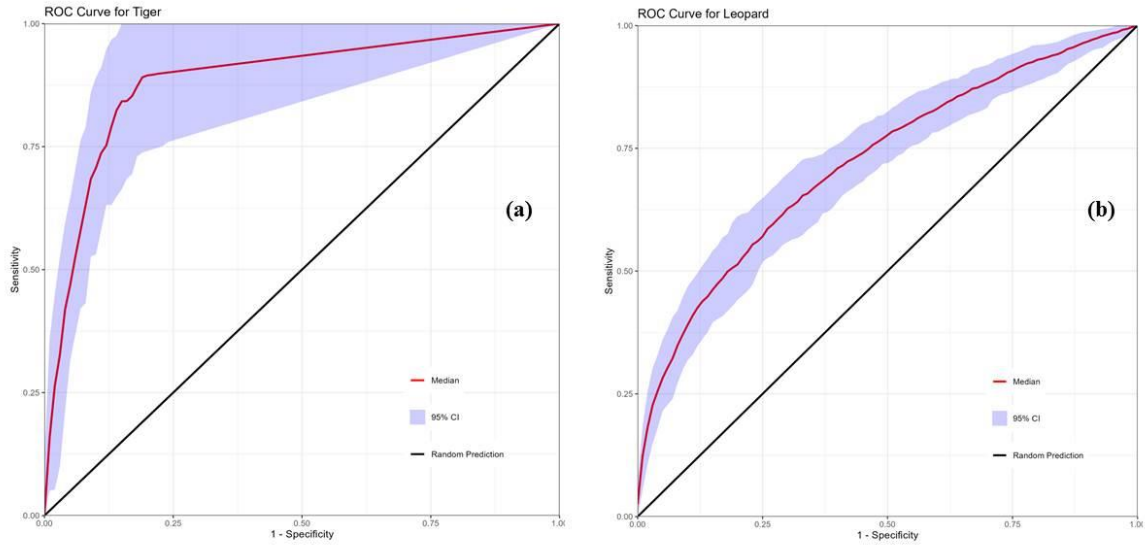

Fig. 3: Receiver Operating Characteristic curves for (a) Tiger (b) Leopard.

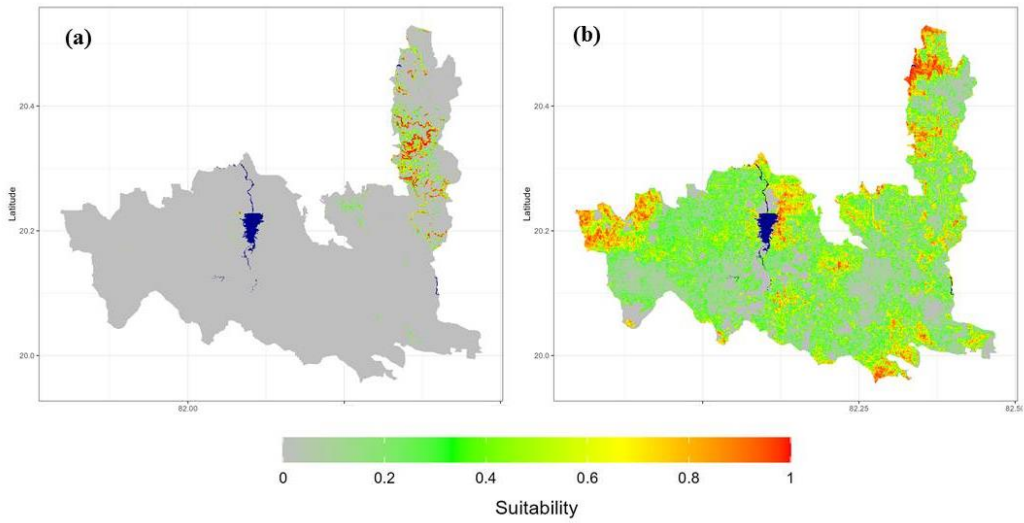

Fig. 4: The predicted habitat suitability of (a) Tiger, and (b) Leopard in USTR.

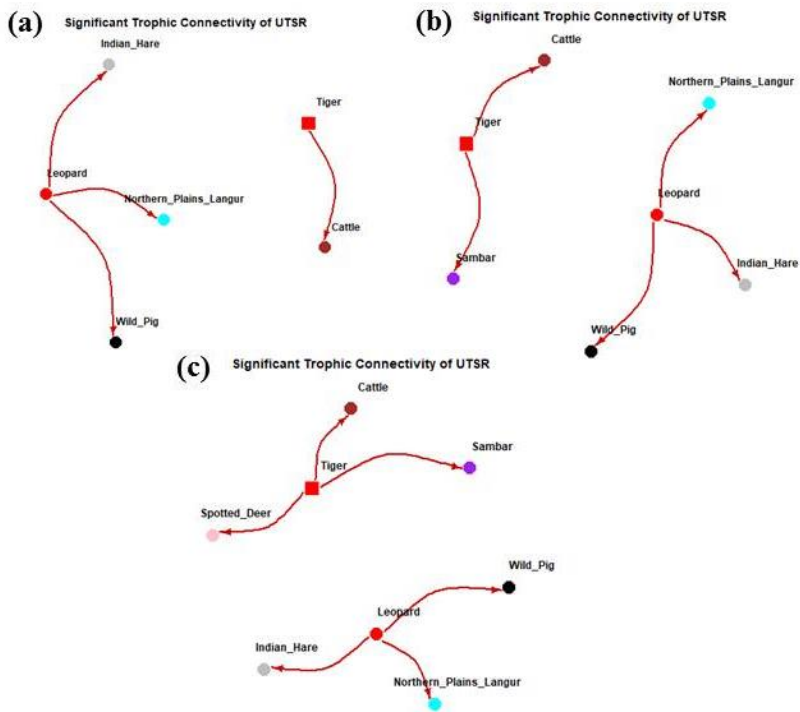

Fig. 5: The significant trophic connection between predators and prey at UTSR, at 90% confidence level (a), at 80% confidence level (b), and at 70% confidence level (c).

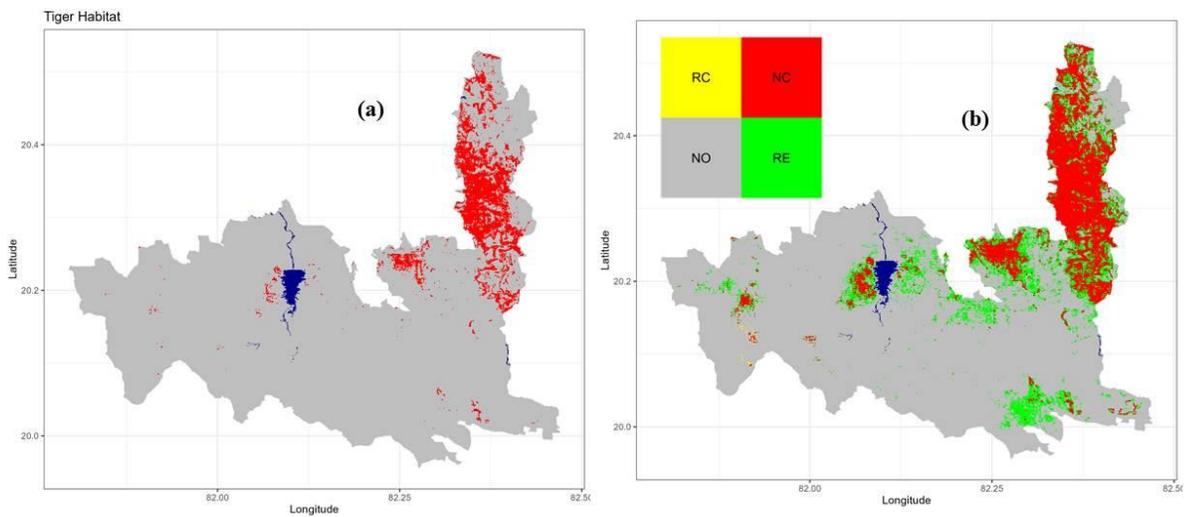

Fig. 6: The realized niche of Tiger (a) and Fundamental niche of Tiger excluding cattle and available Sambar and Spotted deer (b).

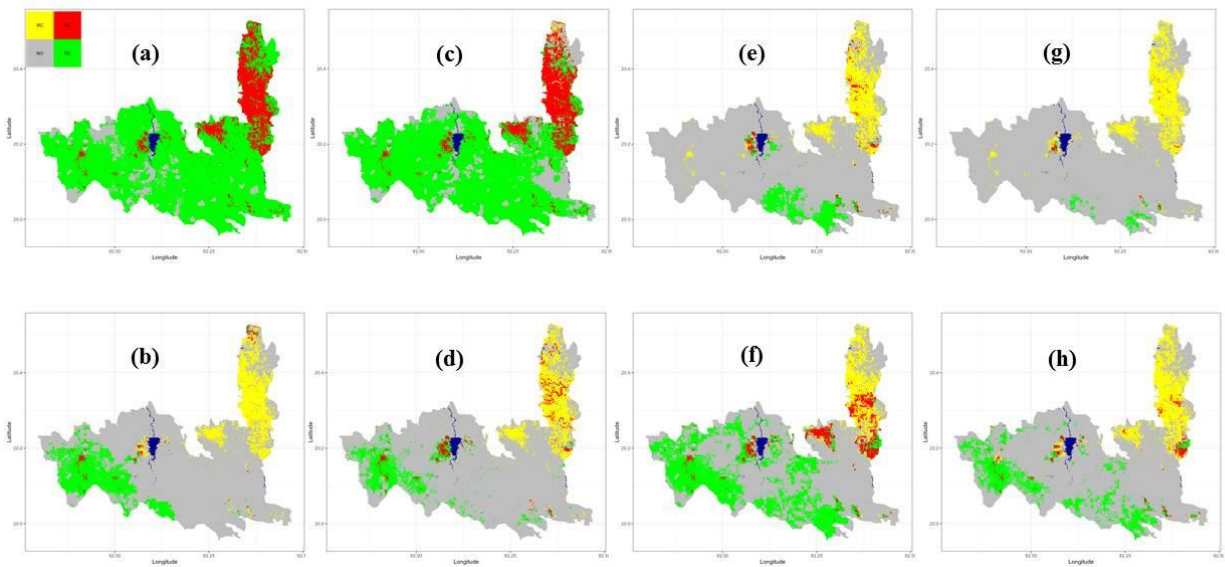

Fig. 7: The projected habitat change for the future scenario 2021-2040, for scenario CMIP6 SSP245 (row 1) and SSP585 (row 2) for models CMCC-ESM2 (column 1), GISS-E2 (column 2), HadGEM3-GC3 (column 3), and UKSM1 (column 4): (a) CMIP6 SSP245 for model CMCC-ESM2, (b) CMIP6 SSP585 for model CMCC-ESM2, (c) CMIP6 SSP245 for model GISS-E2, (d) CMIP6 SSP585 for model GISS-E2, (e) CMIP6 SSP245 for model HadGEM3-GC3, (f) CMIP6 SSP585 for model HadGEM3-GC3, (g) CMIP6 SSP245 for model UKSM1, (h) CMIP6 SSP585 for model UKSM1.

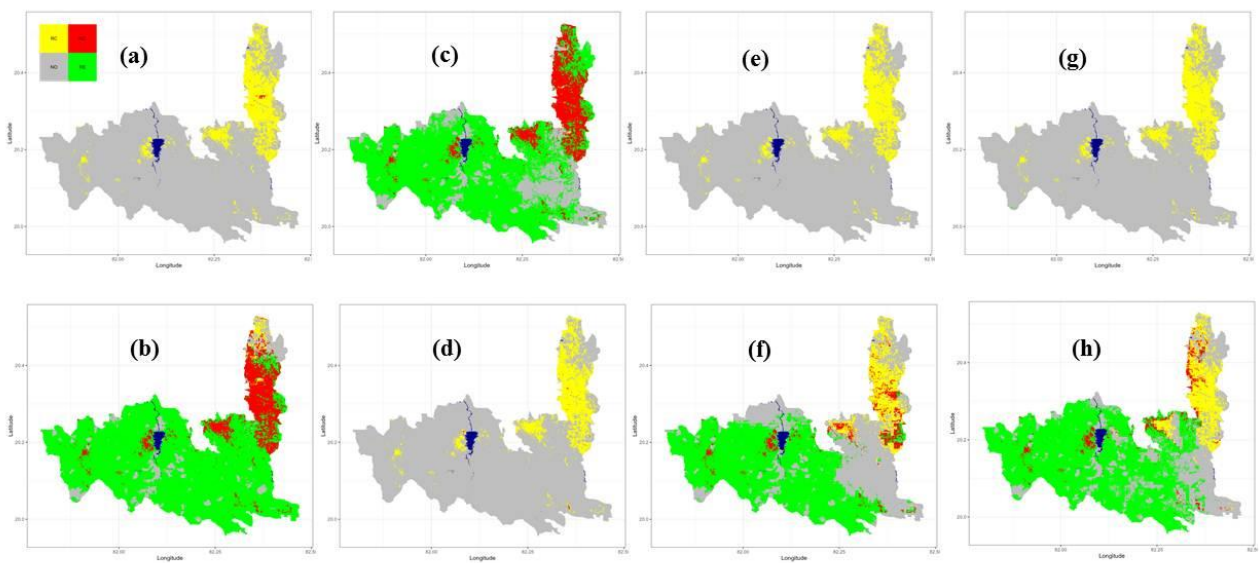

Fig. 8: The projected habitat change for the future scenario 2081-2100, for scenario CMIP6 SSP245 (row 1) and SSP585 (row 2) for models CMCC-ESM2 (column 1), GISS-E2 (column 2), HadGEM3-GC3 (column 3), and UKSM1 (column 4): (a) CMIP6 SSP245 for model CMCC-

ESM2, (b) CMIP6 SSP585 for model CMCC-ESM2, (c) CMIP6 SSP245 for model GISS-E2, (d) CMIP6 SSP585 for model GISS-E2, (e) CMIP6 SSP245 for model HadGEM3-GC3, (f) CMIP6 SSP585 for model HadGEM3-GC3, (g) CMIP6 SSP245 for model UKSM1, (h) CMIP6 SSP585 for model UKSM1.

**Table 1. The 32 environmental factors used in the construction of the Trophic SDM model (including bioclimate, human disturbance, topography and vegetation).**

| Short Name | Description | Source |
|---|---|---|
| **Min_Snare_Distance** | Minimum distance to snare within a 0.1 km grid | Field data, GIS |
| **River_Distances** | Distance to nearest river within a 0.1 km grid | OSM, GIS |
| **Road_Distances** | Distance to nearest road within a 0.1 km grid | |
| **Village_Distances** | Distance to nearest village within a 0.1 km grid | |
| **ASPECT_cat_1** | Fraction of aspect category 1 (e.g., North-facing slopes) | DEM, GIS |
| **ASPECT_cat_2** | Fraction of aspect category 2 (e.g., East-facing slopes) | |
| **ASPECT_cat_3** | Fraction of aspect category 3 (e.g., South-facing slopes) | |
| **ASPECT_cat_4** | Fraction of aspect category 4 (e.g., West-facing slopes) | |
| **DEM_COP30** | Digital Elevation Model (DEM) from COPERNICUS | DEM |
| **ESRI_LULC_cat_11** | Land Use Land Cover category 11 (forests) | ESRI |
| **ESRI_LULC_cat_2** | Land Use Land Cover category 2 (rangelands/cropland) | |
| **SLOPE_cat_1** | Slope category 1 (e.g., gentle slopes, <5 degree) | DEM, GIS |
| **SLOPE_cat_2** | Slope category 2 (e.g., moderate slopes, 5-15 degree) | |
| **SLOPE_cat_3** | Slope category 3 (e.g., steep slopes, 15-30 degree) | |
| **SLOPE_cat_4** | Slope category 4 (e.g., very steep slopes, >30 degree) | |
| **wc2.1_30s_bio_1** | Bioclimatic variable 1 (Annual Mean Temperature) | WorldClim Data |

| | |
|---|---|
| **wc2.1_30s_bio_10** | Bioclimatic variable 10 (Mean Temperature of Warmest Quarter) |
| **wc2.1_30s_bio_11** | Bioclimatic variable 11 (Mean Temperature of Coldest Quarter) |
| **wc2.1_30s_bio_12** | Bioclimatic variable 12 (Annual Precipitation) |
| **wc2.1_30s_bio_13** | Bioclimatic variable 13 (Precipitation of Wettest Month) |
| **wc2.1_30s_bio_14** | Bioclimatic variable 14 (Precipitation of Driest Month) |
| **wc2.1_30s_bio_15** | Bioclimatic variable 15 (Precipitation Seasonality) |
| **wc2.1_30s_bio_16** | Bioclimatic variable 16 (Precipitation of Wettest Quarter) |
| **wc2.1_30s_bio_17** | Bioclimatic variable 17 (Precipitation of Driest Quarter) |
| **wc2.1_30s_bio_18** | Bioclimatic variable 18 (Precipitation of Warmest Quarter) |
| **wc2.1_30s_bio_2** | Bioclimatic variable 2 (Mean Diurnal Range) |
| **wc2.1_30s_bio_3** | Bioclimatic variable 3 (Isothermality) |
| **wc2.1_30s_bio_4** | Bioclimatic variable 4 (Temperature Seasonality) |
| **wc2.1_30s_bio_5** | Bioclimatic variable 5 (Max Temperature of Warmest Month) |
| **wc2.1_30s_bio_6** | Bioclimatic variable 6 (Min Temperature of Coldest Month) |
| **wc2.1_30s_bio_7** | Bioclimatic variable 7 (Temperature Annual Range) |
| **wc2.1_30s_bio_8** | Bioclimatic variable 8 (Mean Temperature of Wettest |

Table 2. Contribution of Environmental variables and selected prey species in suitable habitats of tiger and leopard in USTR. Min_Snare_Distance (Minimum distance to snare within a 0.1 km grid), ASPECT_cat_1 (North facing slopes),ASPECT_cat_2 (East facing slope), ASPECT cat 4 (West facing slope), wc2.1_30s_bio_5 (Max Temperature of Warmest Month), wc2.1_30s_bio_8 (Mean Temperature of Wettest Quarter), wc2.1_30s_bio_11 (Mean Temperature of Coldest Quarter), wc2.1_30s_bio_12 (Annual precipitation), wc2.1_30s_bio_13 (Precipitation for wettest month), ESRI_LULC_cat_11 (Forests), ESRI_LULC_cat_2 (rangelands/croplands), Slope cat 3 (steep slopes, 15-30 degree), Slope cat 4 (very steep slopes, >30 degree), Cattle, Wild Pig, Northern Plains Langur and Indian Hare.

| | Tiger | | |
|---|---|---|---|
| **Variable** | **mean** | **5%** | **95%** |
| **Intercept** | **-9.452** | **-14.666** | **-4.595** |
| Min_Snare_Distance | -1.480 | -2.483 | -0.614 |
| ASPECT_cat_2 | 0.365 | 0.027 | 0.758 |
| SLOPE_cat_4 | 0.799 | 0.267 | 1.355 |
| wc2.1_30s_bio_12 | -5.491 | -10.461 | -0.563 |
| wc2.1_30s_bio_13 | 5.464 | 0.498 | 11.499 |
| Cattle | 2.601 | 1.097 | 4.219 |
| | **Leopard** | | |
| **Variable** | **mean** | **5%** | **95%** |
| **Intercept** | **-2.868** | **-4.858** | **-0.849** |
| ASPECT_cat_1 | 0.151 | 0.057 | 0.250 |
| ASPECT_cat_4 | -0.124 | -0.222 | -0.034 |
| ESRI_LULC_cat_11 | 6.134 | 2.650 | 9.404 |
| ESRI_LULC_cat_2 | 6.226 | 2.694 | 9.564 |
| SLOPE_cat_3 | 0.163 | 0.018 | 0.320 |
| wc2.1_30s_bio_11 | 6.362 | 1.248 | 11.791 |
| wc2.1_30s_bio_12 | 1.284 | 0.165 | 2.372 |
| wc2.1_30s_bio_5 | 2.278 | 0.204 | 4.633 |
| wc2.1_30s_bio_8 | -3.890 | -7.738 | -0.243 |
| Wild_Pig | 0.538 | 0.010 | 1.087 |
| Northern_Plains_Langur | 1.046 | 0.550 | 1.575 |
| Indian_Hare | 0.796 | 0.269 | 1.351 |

Table 3: Table presents habitat changes across different climate change scenarios for the periods 2021-2040 and 2081-2100, using models GISS-E2, HadGEM3-GC3, and UKESM1 under SSP245

and SSP585 scenarios. The categories include No Occurrence (NO), Habitat Contraction (RC), Habitat Expansion (RE), and No Change (NC).

| Experiment | NO | RC | RE | NC |
| --- | --- | --- | --- | --- |
| CMCC-ESM2_ssp245_2021-2040 | 187.52 | 0 | 1519.3 | 255.88 |
| CMCC-ESM2_ssp245_2081-2100 | 1706.82 | 254.25 | 0 | 1.63 |
| CMCC-ESM2_ssp585_2021-2040 | 1433.64 | 238.96 | 273.18 | 16.92 |
| CMCC-ESM2_ssp585_2081-2100 | 183.74 | 21.54 | 1523.08 | 234.34 |
| GISS-E2-1-G_ssp245_2021-2040 | 367.78 | 7.17 | 1339.04 | 248.71 |
| GISS-E2-1-G_ssp245_2081-2100 | 400.74 | 1.69 | 1306.08 | 254.19 |
| GISS-E2-1-G_ssp585_2021-2040 | 1532.88 | 206.8 | 173.94 | 49.08 |
| GISS-E2-1-G_ssp585_2081-2100 | 1706.63 | 254.32 | 0.19 | 1.56 |
| HadGEM3-GC31-LL_ssp245_2021-2040 | 1565.47 | 230.43 | 141.35 | 25.45 |
| HadGEM3-GC31-LL_ssp245_2081-2100 | 1706.82 | 255.88 | 0 | 0 |
| HadGEM3-GC31-LL_ssp585_2021-2040 | 1177.78 | 157.62 | 529.04 | 98.26 |
| HadGEM3-GC31-LL_ssp585_2081-2100 | 703.52 | 177.93 | 1003.3 | 77.95 |
| UKESM1-0-LL_ssp245_2021-2040 | 1677.19 | 247.69 | 29.63 | 8.19 |
| UKESM1-0-LL_ssp245_2081-2100 | 1706.53 | 255.78 | 0.29 | 0.1 |
| UKESM1-0-LL_ssp585_2021-2040 | 1453.21 | 222.45 | 253.61 | 33.43 |
| UKESM1-0-LL_ssp585_2081-2100 | 571.05 | 195.45 | 1135.77 | 60.43 |